\documentclass{article}
\usepackage[utf8]{inputenc}
\usepackage[english]{babel}
\usepackage[T1]{fontenc}
\usepackage{amsmath}
\usepackage{hyperref}
\usepackage{graphicx}
\usepackage{amssymb}
\usepackage{xfrac}
\usepackage{multirow}

\renewcommand{\d}[1]{\mathrm{d}{#1}}

\newcommand\numberthis{\addtocounter{equation}{1}\tag{\theequation}}

\title{General framework for the analysis of imperfections in nonlinear systems.}
\author{Matteo Santandrea$^{1,*}$ \and Michael Stefszky$^1$ \and Christine Silberhorn$^1$}
\date{$^1$ Integrated Quantum Optics, Paderborn University, Warburger Stra{\ss}e 100, 33098 Paderborn, Germany\\$^*$ Corresponding author: matteo.santandrea@upb.de}

\begin{document}
\maketitle
\begin{abstract}
In this paper, we derive a framework to understand the effect of imperfections on the phasematching spectrum of a wide class of nonlinear systems.
We show that this framework is applicable to many physical systems, such as waveguides or fibres.
Furthermore, this treatment reveals that the product of the system length and the magnitude of the imperfections completely determines the phasematching properties of these systems, thus offering a general rule for system design.
Additionally, our framework provides a simple method to compare the performance of a wide range of nonlinear systems.
\end{abstract}

\section{Introduction}
Both in classical and in quantum optics, nonlinear phasematching processes are fundamental tools for the generation, manipulation and detection of a plethora of different states of light, e.g. frequency doubling \cite{Franken1961}, pulse spectra characterisation \cite{Kane1993, Iaconis1998}, parametric downconversion for photon pair generation \cite{Harder2013}, frequency up-conversion for enhanced single photon detection \cite{Pelc2011b} and frequency conversion for interfacing different quantum memories \cite{Maring2017}.
These processes are usually realised in $\chi^{(2)}$ or $\chi^{(3)}$ nonlinear systems, e.g. lithium niobate crystals (bulk or waveguides) or photonic crystal fibers (PCFs). 
The fabrication of such systems, despite in many cases being very mature, is still affected by imperfections that spoil the phasematching spectrum of the process. 
This spectrum is the critical parameter in the case of many quantum system, such as the quantum pulse gate \cite{Eckstein2011}; furthermore the quality of the spectrum is directly related to the efficiency \cite{Santandrea2019}.  
Therefore, in the past decades, several studies have discussed the relation between fabrication errors in waveguides \cite{Lim1990, Helmfrid1991, Helmfrid1992, Pelc2010, Pelc2011a, Phillips2013, Nouroozi2017, Santandrea2019} and in fibres \cite{Karlsson1998, Farahmand2004, FrancisJones2016, Harle2017} and their spectral performance.


Previous investigations have typically considered only specific types of imperfections.  
However, comparing all these studies, one can note striking similarities among the presented results. 
Indeed, all the systems analysed exhibit a close connection between the device length, the amount of imperfections and the overall performance of the nonlinear process. 
This observation suggests the existence of a scaling law, common to all nonlinear systems, determining the length where the process becomes sensitive to the imperfections present in the system. 

In this letter we show that such a scaling law indeed exists and derive a general, system-independent framework to understand the effect of imperfections on the performance of a wide class of nonlinear systems. 
We show that our framework can provide important rules to predict and design the behaviour of many nonlinear systems. 

\section{Mathematical formulation}
\label{sec:mathematical_formulation}
Consider a nonlinear process in a system of length $L$ characterized by a momentum mismatch 
\begin{equation}
\Delta\beta = \sum_i s_i\beta_i=\sum_i s_i \frac{2\pi n_i(\lambda_i)}{\lambda_i},
\label{eq:momentum_mismatch}
\end{equation} 
where $\beta_i$, $n_i$ and $\lambda_i$ are the propagation constant, the refractive index and the wavelength of the $i$-th field, and $s_i=\pm1$ is a sign that depends on the type of process considered; for example, for copropagating three wave mixing $\Delta\beta = \beta_3-\beta_2-\beta_1$.
Note that Eq.~(\ref{eq:momentum_mismatch}) is valid for any general wave mixing process.
The phasematching spectrum of the nonlinear process, normalized per unit length, is defined as \cite{Helmfrid1991} 
\begin{equation}
\Phi = \frac{1}{L}\int_0^L \exp\left\lbrace\mathrm{i} \int_0^z\Delta\beta(\xi)\d{\xi}\right\rbrace \d{z},\label{eq:generalized_pm_integral}
\end{equation}
where $z$ denotes the propagation axis along the system and scaling constants have been neglected since they do not affect the shape of the phasematching spectrum. 
Note that Eq.~(\ref{eq:generalized_pm_integral}) sets the ideal maximum efficiency is 1. 
Typically, the phasematching $\Phi$ is expressed as a function of the wavelengths or the frequencies of the fields involved in the process. This however prevents a direct comparison of different systems, since $\Delta\beta$ depends nonlinearly on these parameters, as shown in \eqref{eq:momentum_mismatch}. Therefore, in the rest of the paper, we will consider the phasematching as a function of the $\Delta\beta$.

Under ideal fabrication and operation conditions, the momentum mismatch $\Delta\beta$ is constant along the sample.
However, fabrication imperfections and/or non-ideal operating conditions affect the phase mismatch of the process and they can be described as a position-dependent $\Delta\beta(z)$.
If the variation of the momentum mismatch is sufficiently small such that it can be considered frequency-independent \cite{Chang2014}, we can introduce the \textit{decoupling approximation}
\begin{equation}
\Delta\beta(z)\approx \Delta\beta_0 + \sigma\delta\beta(z),\label{eq:decoupling_approx}
\end{equation}
where the momentum mismatch has been decomposed into the sum of $\Delta\beta_0$, describing the momentum mismatch of the process in absence of inhomogenenities, and $\sigma\delta\beta(z)$, that encompasses the variation of $\Delta\beta$ due to inhomogeneities in the system. 
The noise amplitude $\sigma$ is chosen such that $\left|\delta\beta(z)\right|\leq 1$.

Under these assumptions and with a change of variables $\sfrac{z}{L}\rightarrow z'$ and $\sfrac{\xi}{L}\rightarrow \xi'$, the integral in (\ref{eq:generalized_pm_integral}) can be rewritten as:
\begin{equation}
\Phi(\Delta\beta_0 L) = \int_0^1 \exp\left\lbrace\mathrm{i}\Delta\beta_0 L z'\right\rbrace\times
\exp\left\lbrace\mathrm{i}\sigma L \int_0^{z'} \delta\beta(L\xi')\d{\xi'}\right\rbrace\d{z'},\label{eq:reworked_integral}
\end{equation}
where the first exponential term leads to the usual sinc dependence of the phasematching $\Phi$ on the mismatch $\Delta\beta_0$, while the second exponential term describes the effect of the noise $\sigma\delta\beta(z)$ on the system.
Eq.~(\ref{eq:reworked_integral}) can also be understood as the Fourier transform of the rectangular function representing the crystal, multiplied by a phase factor introduced by the imperfections.

In particular, the first exponential shows that the phasematching spectrum of all noiseless systems is identical, bar a scaling factor given by the length of the system. 
The second exponential highlights that all systems with the same \textit{noise-length product} $\sigma L$ and noise profile $\delta\beta(z)$, defined for $z\in[0,L]$, will exhibit the same phasematching spectrum. 
This allows us to study the effect of variations of the momentum mismatch on a system with unit length and then extrapolate the results to systems with any length, provided the correct scaling $\Delta\beta_0 \rightarrow \Delta\beta_0 L$ and $\sigma\rightarrow \sigma L$ is applied.

\section{Simulation of inhomogeneous systems}
\label{sec:simulations}
In the previous section it was shown that the phasematching spectrum $\Phi(\Delta\beta_0)$ is fully characterized by the noise-length product $\sigma L$ and the noise profile $\delta\beta(z)$.
Therefore, in the following we study the impact of these two parameters on the profile of the phasematching spectrum.

The scaling law presented in the previous section allows us to consider a general nonlinear system with $L=1$m and $\sigma\in$ [0.001, 1000] m$^{-1}$ without loss of generality.
We model $\delta\beta(z)$ as a stochastic process with a $1/f$ spectral density to describe the long range correlations that can arise due to fabrication imperfections and/or under non-ideal operating conditions of the nonlinear system \cite{Santandrea2019}.
For each value of $\sigma L$, we randomly generate 100 different $\delta\beta(z)$ profiles and calculate the relative phasematching as a function of $\Delta\beta_0$ using a piecewise approximation \cite{FrancisJones2016} of Eq.~(\ref{eq:reworked_integral}):
\begin{multline}
\Phi(\Delta\beta_0) = \sum_{n=1}^{N}\mathrm{sinc}\left( \frac{\Delta\beta_n \Delta z_n}{2}\right)\exp\left\lbrace\mathrm{i} \frac{\Delta\beta_n\Delta z_n}{2}\right\rbrace \times\\ \exp\left\lbrace\mathrm{i}\sum_{m=1}^{n-1}\Delta\beta_m \Delta z_m\right\rbrace,
\end{multline}
where $\Delta z$ is the mesh discretisation along the $z$ axis, such that $\sum_n\Delta z_n =L$, and $\Delta\beta_n = \Delta\beta_0 + \delta\beta(z_n)$.

To quantify the difference between the phasematching $\Phi_{noisy}$ of a system with imperfections and the phasematching $\Phi_{ideal}$ of an ideal one, i.e. where $\delta\beta(z) = 0$, we introduce the fidelity $\mathcal{F}$, defined as 
\begin{align}
\mathcal{F} = \frac{\max_{\tau} \int_{-\infty}^{+\infty} I_{ideal}(\Delta\beta_0) I_{noisy}(\Delta\beta_0-\tau)\d{\Delta\beta_0}}
{\int_{-\infty}^{+\infty}I_{ideal}^2(\Delta\beta_0)\d{\Delta\beta_0}},
\label{eq:fidelity}
\end{align}
where $I = \left|\Phi\right|^2$. 
In Eq.~(\ref{eq:fidelity}), the two curves are normalized such that $\int \left|\Phi_{noisy}\right|^2\d{\Delta\beta} = \int \left|\Phi_{ideal}\right|^2\d{\Delta\beta}$, since this quantity is conserved in the presence of momentum mismatch variation \cite{Nash1970}.
Using this definition, the fidelity approaches 1 if the effect of noise on the phasematching spectrum is negligible and tends to 0 if the contribution of the noise is dominant.

\begin{figure}[btp]
\centering
\includegraphics[width = 0.8\columnwidth]{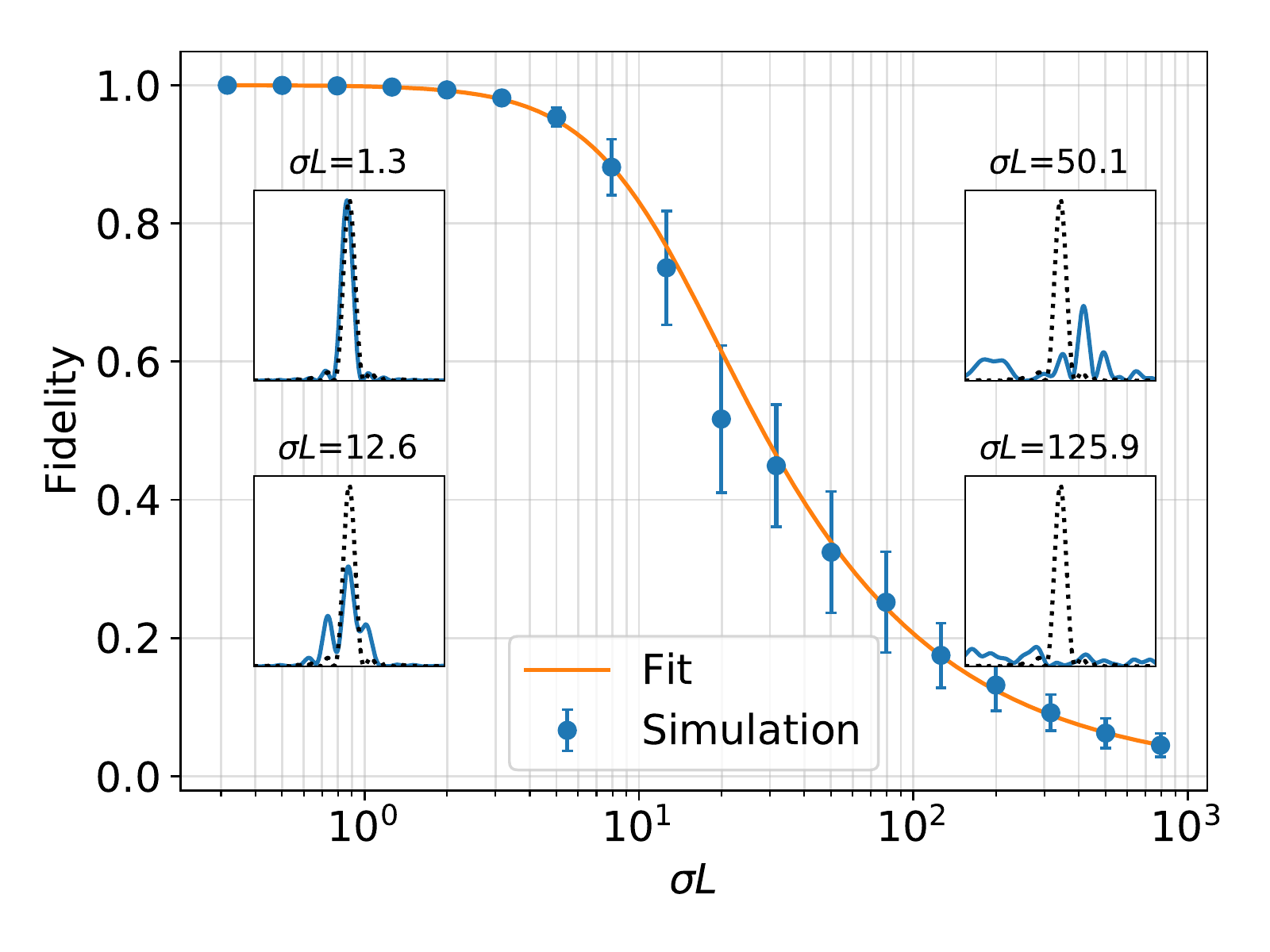}
\caption{Simulated reduction in fidelity $\mathcal{F}$ as the noise-length product $\sigma L$ of a nonlinear system increases.  
Errorbars indicate the standard deviations calculated from 100 randomly generated samples.
The solid orange line corresponds to the best fit of the average fidelity and follows the relation given by Eq.~(\ref{eq:model_fit}). 
The insets show examples of simulated phasematching spectra (solid blue line) compared to the ideal phasematching spectra (dotted black line), for chosen $\sigma L$ values.}\label{img:sim_results}
\end{figure}

We calculate the fidelity $\mathcal{F}$ for the simulated phasematching spectra and the results are plotted in Figure \ref{img:sim_results}. 
The results can be well approximated by a Lorentzian-like fitting curve, shown in Figure \ref{img:sim_results} with a solid orange line:
\begin{equation}
\mathcal{F}(\sigma L) = \frac{1}{\left[1+A \cdot(\sigma L)^B\right]^C},
\label{eq:model_fit}
\end{equation}
with $A=5.4(3)\times 10^{-3}$, $B=2.12(4)$, $C=0.35(2)$.

The simulations show that systems with  $\sigma L\leq 10$ have a fidelity close to 1, while for $\sigma L> 10$ the average fidelity rapidly drops below 0.5. 
Therefore, the condition $\sigma L \leq10$ represents a general design principle for these systems.


\section{General design rule for nonlinear systems}
We now move away from the abstract description, in terms of $\delta\beta$, to study how fabrication imperfections directly relate to the phasematching and to show how the condition $\sigma L\leq 10$ aids in designing a given nonlinear process.
For simplicity, we assume that all the imperfections are introduced by a single system parameter $f$. 
For example, $f$ could represent the local temperature of the system during operations, the width of a waveguide or the holes' diameter in a PCF.
With a suitable model of the system, one can relate the noise amplitude $\sigma$ to the variation of the parameter $f$ with a Taylor expansion $\sigma \approx |\partial_f\Delta\beta| \delta f$. 
Therefore, the condition $\sigma L\leq 10$ can be rewritten as
\begin{equation}
\sigma L\leq 10 \Rightarrow \delta f\cdot L\leq \frac{10}{\left|\partial_f\Delta\beta\right|}.
\label{eq:fabrication_inequality}
\end{equation}
In this form, the trade-off between the physical parameters characterizing the sample, namely its length $L$ and the error $\delta f$, is explicitly revealed.

If $\left|\partial_f\Delta\beta\right|$ is known, with the help of Eq.~(\ref{eq:fabrication_inequality}) one can bound the maximum length of the system to the maximum error during fabrication/operation in order to ensure high fidelity. 
This can provide crucial information during the design of samples and experiments: if the error $\delta f$ cannot be further reduced, then the maximum length of the system to achieve high fidelity is bounded by (\ref{eq:fabrication_inequality}); viceversa, if the length of the sample is constrained by the experiment, then the error $\delta f$ has to be minimized to satisfy (\ref{eq:fabrication_inequality}).

As an example, we consider the restraints set by Eq.~(\ref{eq:fabrication_inequality}) on the four wave mixing, seeded parametric downconversion process in a PCF described in \cite{FrancisJones2016}. 
In the paper, the authors show that a 3m-long fibre presents a very distorted phasematching, while a 15-cm long piece of the same fibre is characterized by a much cleaner but still imperfect spectrum. 
In particular, they investigate the effects of the variation of the pitch $\Lambda$  of the holes and their diameter $d$ around the ideal design parameters $\Lambda_0 = 1.49\mu$m and $d=0.6414\mu$m. 
Using the Sellmeier equations provided in \cite{Saitoh2005}, we can estimate the effect of the variation of these parameters by calculating the partial derivatives 
\begin{align*}
\left|\partial_\Lambda\Delta\beta\right|_{\Lambda_0, d_0}&\approx 2\times 10^{-4}\mu\mathrm{m}^{-2}\\
\left|\partial_d\Delta\beta\right|_{\Lambda_0, d_0} &\approx 1.5\times 10^{-2}\mu\mathrm{m}^{-2} \numberthis\label{eq:delta_beta_pcf}
\end{align*}
Since $\partial_d\Delta\beta$ is two orders of magnitude higher than $\partial_\Lambda\Delta\beta$, the resulting phasematching is much more sensitive to variations of the holes' diameter $d$ rather than to variations in the pitch. 

From the observation that the reported phasematching is already degraded for PCF longer than 30cm  (which implies that $\sigma L \geq 10$); 
from this, we can infer that the original 3m-long PCF had a $\sigma L\geq 100$.
Using Eq.~(\ref{eq:model_fit}) the expected fidelity for this noise-length product is below $0.2$, thereby explaining the distorted phasematching spectrum measured in \cite{FrancisJones2016}.
Finally, combining (\ref{eq:fabrication_inequality}) and  (\ref{eq:delta_beta_pcf}), we can estimate that it is necessary to limit $\delta\Lambda$ ($\delta d$) below 1.1\% (0.078\%) to achieve high fidelity phasematching in a 3m-long fibre, clearly a challenging task.

\section{Comparison with simulations of different physical systems}
\label{sec:comparison}
The \textit{decoupling approximation} introduced in Eq.~(\ref{eq:decoupling_approx}) relies on the assumption that the refractive index variation due to imperfections can be considered independent of the wavelength \cite{Chang2014}.
To show that this approximation is indeed valid in many cases of interest, we now compare the results presented in Figure \ref{img:sim_results} with simulations of a number of systems affected by different sources of imperfections, all presenting a $1/f$ noise spectrum.

The investigated systems are: (a) type-0 second harmonic generation (SHG) in titanium in-diffused lithium niobate (Ti:LN) channel waveguides, with errors $\delta w$ in the waveguide width;
(b) type-II SHG in rubidium exchanged potassium titanyl phosphate (Rb:KTP) channel waveguides, with errors $\delta w$ in the waveguide width; 
(c) type-II sum frequency generation (SFG) in a bulk LN crystal, with an inhomogeneous temperature profile with maximum excursion $\delta T$;
(d) four wave mixing SFG in a PCF, with errors $\delta d$ in the holes' diameter.
The lengths of the simulated devices and the fabrication errors are reported in Table \ref{table:parameters}, while the details of the simulations are presented in the appendix.
For each system (a-d), we calculate the phasematching and the fidelity $\mathcal{F}$ of 20 randomly generated samples for every combination of the parameters in Table \ref{table:parameters}.

\begin{figure}[btp]
\centering
\includegraphics[width = 0.8\columnwidth]{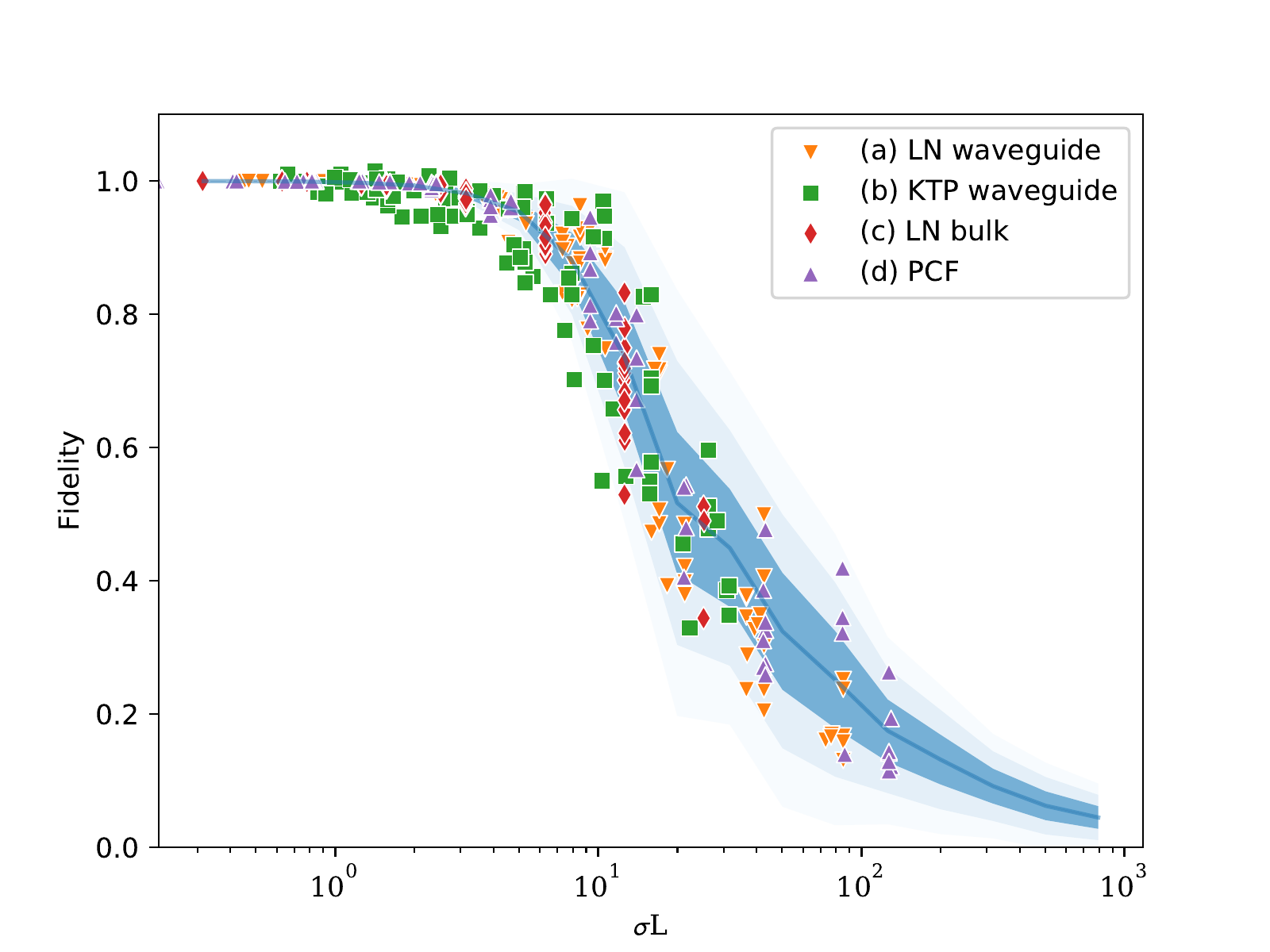}
\caption{Fidelity $\mathcal{F}$ simulated for different real systems in the presence of $1/f$ noise on a fabrication/operation parameter. The solid blue line represents the average fidelity as estimated by the general model, while the shaded areas correspond to 1,2 and 3 standard deviations. Details about the processes are provided in the main text and in the appendix.}\label{img:systemsim}
\end{figure}

To aid in visualisation, a randomly chosen subset of the calculated values of $\mathcal{F}$ is presented in Figure \ref{img:systemsim}.
It is apparent that the fidelity of the simulated processes closely follows the model derived in the previous section, despite having different noise sources and being realized in vastly disparate physical systems.
This shows that the model presented provides a general framework to analyse the effects of inhomogeneities on the phasematching performance of a wide range of nonlinear systems.

\begin{table}[tbp]
\centering
\begin{tabular}{| l | l |}
\hline
\multirow{2}{*}{(a)} & $L$ = 5, 10, 40, 80 mm\\
&	$\delta w$ = 0.05, 0.1, 0.25, 0.5 $\mu$m\\
\hline
\multirow{2}{*}{(b)} & $L$ = 5, 10, 15, 20, 25, 30 mm\\
&	$\delta w$ = 0.05, 0.1, 0.2, 0.3, 0.5 $\mu$m\\
\hline
\multirow{2}{*}{(c)} & $L$ = 5, 10, 20, 40 mm\\
&	$\delta T$ = 0.1, 0.2, 0.5, 1.0, 2.0 $^\circ$C\\
\hline
\multirow{2}{*}{(d)} & $L$ = 0.5, 1, 2, 3 m\\
&	$\delta d$ = 0.64, 6.41, 64.14,   641.44 pm\\
\hline
\end{tabular} 
\caption{Fabrication parameters used for the simulation of (a) Ti:LN waveguides, (b) channel Rb:KTP waveguide, (c) ridge Rb:KTP waveguide and (d) photonic crystal fibre. Details about the processes are provided in the main text and in the appendix.}
\label{table:parameters}
\end{table}

\section{Conclusion}
In this paper, a general framework for the description and understanding of the phasematching of nonlinear processes in the presence of momentum mismatch variations has been developed.
In particular, we have shown that the shape of the phasematching spectrum of a wide class of nonlinear systems is uniquely determined by the \textit{noise-length product} $\sigma L$ and the noise profile $\delta\beta(z)$.
This result shows that it is possible to study the effect of variations of $\Delta\beta$ independent of the specific physical properties of the nonlinear systems and of the sources of imperfections. 

Using this framework, we investigated the effect that a noise profile $\delta\beta(z)$ with a $1/f$ noise spectrum has on the phasematching spectrum, for different noise-length products.
We introduced a process fidelity $\mathcal{F}$ to measure the ideality of a phasematching spectrum and discovered that high fidelities ($\mathcal{F}>0.8$) are found for systems with $\sigma L$<10.
This inequality provides a general design rule for realising high-fidelity nonlinear processes.

We applied this design rule to analyse the case of fabrication errors in the photonic crystal fibre reported in \cite{FrancisJones2016}. 
The analysis was able to explain the reported phasematching spectra and provide insight towards the requirements necessary to achieve a high-fidelity phasematching spectrum.

Finally, we show that many different physical systems follow the trend predicted by the model. 
This shows that the presented framework provides a universal method to understand and compare the properties of a wide range of nonlinear processes.

\section{Acknowledgement}
The work was supported by the European Union via the EU quantum flagship project UNIQORN (Grant No. 820474) and by the DFG (Deutsche Forschungsgemeinschaft).
The authors thank Gana\"{e}l Roeland, Vahid Ansari, Jeremy Kelly-Massicotte and Pete Mosley for helpful discussions.

\appendix
\section{Models for the nonlinear system investigated in sec. \ref{sec:comparison}}
In section \ref{sec:comparison} we simulated four different nonlinear systems with imperfections that change the momentum mismatch along the propagation axis. 
We outline here in detail the models employed for the different simulated systems.

Process (a) is a type-0 (zzz) second harmonic generation (SHG) 1550nm$\rightarrow$775nm in Z-cut, X-propagating titanium in-diffused lithium niobate (Ti:LN) channel waveguides. Before indiffusion, the Ti stripe thickness is 80nm and its width is 7$\mu$m with a noise which has a maximum excursion given by $\delta w$. We considered a diffusion time of 8.5h and temperature of 1060$^\circ$C. The operation temperature is considered fixed at 190$^\circ$C. The model used to calculate the Sellmeier equations is described in \cite{Strake1988}.

Process (b) is a type-II (yzy) SHG 1550nm$\rightarrow$775nm in Z-cut, X-propagating rubidium exchanged potassium titanyl phosphate (Rb:KTP) channel waveguides. We assumed waveguide widths of 3$\mu$m with a noise which has a maximum excursion of $\delta w$. The diffusion depth is set to 8$\mu$m. We model the dispersion of the waveguide using the model in \cite{Callahan2014}.

Process (c) is a type-II (yzy) sum frequency generation (SFG) 1550nm+875nm$\rightarrow$559nm in a Z-cut, X-propagating bulk LN crystal. We assumed position-dependent temperature profile of the crystal of 190$^\circ$C with a noise which has a maximum excursion of $\delta T$. We used the Sellmeier equation reported in \cite{Jundt1997}.

Process (d) is a sum frequency generation 1545nm+805nm$\rightarrow$1058.5nm in a PCF \cite{FrancisJones2016}, using the model in \cite{Saitoh2005}. The nominal pitch is $\Lambda = 1.49\mu$m and the holes' diameter is 641.4nm with a noise which has a maximum excursion of $\delta d$.


\begin{thebibliography}{10}

\bibitem{Franken1961}
P.~A. Franken, A.~E. Hill, C.~W. Peters, and G.~Weinreich.
\newblock {Generation of optical harmonics}.
\newblock {\em Physical Review Letters}, 7(4):118--119, 1961.

\bibitem{Kane1993}
Daniel~J Kane and Rick Trebino.
\newblock {Characterization of Arbitrary Femtosecond Pulses Using
  Frequency-Resolved Optical Gating}.
\newblock {\em IEEE Journal of Quantum Electronics}, 29(2), 1993.

\bibitem{Iaconis1998}
C~Iaconis and Ian~A. Walmsley.
\newblock {Spectral phase interferometry for direct electric-field
  reconstruction of ultrashort optical pulses}.
\newblock {\em Optics Letters}, 23(10):792--794, 1998.

\bibitem{Harder2013}
Georg Harder, Vahid Ansari, Benjamin Brecht, Thomas Dirmeier, Christoph
  Marquardt, and Christine Silberhorn.
\newblock {An optimized photon pair source for quantum circuits}.
\newblock {\em Optics Express}, 21(12), 2013.

\bibitem{Pelc2011b}
J.~S. Pelc, L.~Ma, C~R Phillips, Qiang Zhang, C~Langrock, O~Slattery, X~Tang,
  Martin~M Fejer, G~N Gol, O~Okunev, G~Chulkova, A~Lipatov, A~Semenov,
  K~Smirnov, B~Voronov, A~Dzardanov, C~Williams, and R~Sobolewski.
\newblock {Long-wavelength-pumped upconversion single-photon detector at 1550
  nm: performance and noise analysis}.
\newblock {\em Optics Express}, 19(22), 2011.

\bibitem{Maring2017}
Nicolas Maring, Pau Farrera, Kutlu Kutluer, Margherita Mazzera, Georg Heinze,
  and Hugues de~Riedmatten.
\newblock {Photonic quantum state transfer between a cold atomic gas and a
  crystal}.
\newblock {\em Nature}, 551(7681):485--488, 2017.

\bibitem{Eckstein2011}
Andreas Eckstein, Benjamin Brecht, and Christine Silberhorn.
\newblock {A quantum pulse gate based on spectrally engineered sum frequency
  generation}.
\newblock {\em Optics Express}, 19(15):13770, 2011.

\bibitem{Santandrea2019}
Matteo Santandrea, Michael Stefszky, Vahid Ansari, and Christine Silberhorn.
\newblock {Fabrication limits of waveguides in nonlinear crystals and their
  impact on quantum optics applications}.
\newblock {\em New Journal of Physics}, 21(033038), 2019.

\bibitem{Lim1990}
E.~J. Lim, S.~Matsumoto, and Martin~M Fejer.
\newblock {Noncritical phase matching for guided-wave frequency conversion}.
\newblock {\em Applied Physics Letters}, 57(22):2294--2296, 1990.

\bibitem{Helmfrid1991}
Sten Helmfrid and Gunnar Arvidsson.
\newblock {Influence of randomly varying domain lengths and nonuniform
  effective index on second-harmonic generation in quasi-phase-matching
  waveguides}.
\newblock {\em Journal of Opt. Soc. Am. B}, 8(4), 1991.

\bibitem{Helmfrid1992}
Sten Helmfrid, Gunnar Arvidsson, and Jonas Webj{\"{o}}rn.
\newblock {Influence of various imperfections on the conversion efficiency of
  second-harmonic generation in quasi-phase-matching lithium niobate
  waveguides}.
\newblock {\em Journal of Opt. Soc. Am. B}, 10(2):222--229, 1992.

\bibitem{Pelc2010}
J.~S. Pelc, C~Langrock, Qiang Zhang, and Martin~M Fejer.
\newblock {Influence of domain disorder on parametric noise in
  quasi-phase-matched quantum frequency converters.}
\newblock {\em Optics letters}, 35(16):2804--2806, 2010.

\bibitem{Pelc2011a}
J.~S. Pelc, C~R Phillips, D~Chang, C~Langrock, and Martin~M Fejer.
\newblock {Efficiency pedestal in quasi-phase-matching devices with random
  duty-cycle errors.}
\newblock {\em Optics letters}, 36(6):864--866, 2011.

\bibitem{Phillips2013}
C~R Phillips, J.~S. Pelc, and Martin~M Fejer.
\newblock {Parametric processes in quasi-phasematching gratings with random
  duty cycle errors}.
\newblock {\em Journal of the Optical Society of America B}, 30(4):982--993,
  2013.

\bibitem{Nouroozi2017}
Rahman Nouroozi.
\newblock {Effect of Waveguide Inhomogeneity in a $\chi$(2)-Based Pulsed
  Optical Parametric Amplifie}.
\newblock {\em Journal of Lightwave Technology}, 35(9):1693--1699, 2017.

\bibitem{Karlsson1998}
Magnus Karlsson.
\newblock {Four-wave mixing in fibers with randomly varying zero-dispersion
  wavelength}.
\newblock {\em Journal of the Optical Society of America B}, 15(8):2269, 1998.

\bibitem{Farahmand2004}
Mitra Farahmand and Martijn de~Sterke.
\newblock {Parametric amplification in presence of dispersion fluctuations.}
\newblock {\em Optics Express}, 12(1):136--142, 2004.

\bibitem{FrancisJones2016}
Robert J.~A. Francis-Jones and Peter~J. Mosley.
\newblock {Characterisation of longitudinal variation in photonic crystal
  fibre}.
\newblock {\em Optics Express}, 24(22):24836, 2016.

\bibitem{Harle2017}
T~Harl{\'{e}}, M~Barbier, M~Cordier, A~Orieux, E~Diamanti, I~Zaquine, and Ph.
  Delaye.
\newblock {Constraints on photonic crystal fibers homogeneity for photon pair
  generation}.
\newblock In {\em Quantum Information and Measurement (QIM) 2017}, page
  QT6A.19. Optical Society of America, 2017.

\bibitem{Chang2014}
Derek Chang, Carsten Langrock, Yu-wei Lin, C~R Phillips, C~V Bennett, and
  Martin~M Fejer.
\newblock {Complex-transfer-function analysis of optical-frequency converters}.
\newblock {\em Optics Letters}, 39(17):5106--5109, 2014.

\bibitem{Nash1970}
F.~R. Nash, G.~D. Boyd, M.~III Sargent, and P.~M. Bridenbaugh.
\newblock {Effect of Optical Inhomogeneities on Phase Matching in Nonlinear
  Crystals}.
\newblock {\em Journal of Applied Physics}, 41(6):2564, 1970.

\bibitem{Saitoh2005}
Kunimasa Saitoh and Masanori Koshiba.
\newblock {Empirical relations for simple design of photonic crystal fibers}.
\newblock {\em Optics Express}, 13(1):267, 2005.

\bibitem{Strake1988}
E.~Strake, G.~P. Bava, and I.~Montrosset.
\newblock {Guided Modes of Ti:LiNbO3 Channel Waveguides: A Novel
  Quasi-Analytical Technique in Comparison with the Scalar Finite-Element
  Method}.
\newblock {\em Journal of Lightwave Technology}, 6(6):1126--1135, 1988.

\bibitem{Callahan2014}
Patrick~T. Callahan, Kemal Shafak, Philip~R. Battle, Tony~D. Roberts, and
  Franz~X. K{\"{a}}rtner.
\newblock {Fiber-coupled balanced optical cross-correlator using PPKTP
  waveguides}.
\newblock {\em Optics Express}, 22(8):1423--1431, 2014.

\bibitem{Jundt1997}
D.H. Jundt.
\newblock {Temperature-dependent Sellmeier equation for the index of
  refraction, n(e), in congruent lithium niobate.}
\newblock {\em Optics Letters}, 22(20):1553--1555, 1997.

\end{thebibliography}

\end{document}